\documentclass[10pt,leqno]{amsart}
\usepackage{graphicx}
\usepackage{indentfirst,csquotes}

\usepackage{pdflscape}
\usepackage{geometry}
\usepackage{amssymb,amsthm,amsmath}
\usepackage{xcolor,paralist,hyperref,titlesec,fancyhdr,etoolbox}

\titleformat{\section}[display]{\normalfont\huge\bfseries\centering}{}{10pt}{\Large}
\titlespacing*{\section}{0pt}{0ex}{0ex}

\hypersetup{ colorlinks=true, linkcolor=black, filecolor=black, urlcolor=black }

\begin{document}

\title{Classifying Tokenised Money: Dimensions and Design Features}
\author[Ankenbrand, Bieri, Ferrazzini, Höhener]{Thomas Ankenbrand\textsuperscript{1}, Denis Bieri\textsuperscript{2}, Stefano Ferrazzini\textsuperscript{3}, Johannes Höhener\textsuperscript{4}}
\date{}
\address{}
\email{}
\maketitle

\begin{abstract}
Tokenised money encompasses a broad range of digital monetary instruments issued on distributed ledger technology, including Central Bank Digital Currencys (CBDCs), deposit tokens, stablecoins, and decentralised protocol-based designs. Despite their shared monetary function, these instruments differ markedly in issuer structure, collateralisation, stability mechanisms, governance, and technological embedding, creating conceptual ambiguity. This paper proposes a concise taxonomy spanning twelve key design dimensions, offering a systematic framework for comparing heterogeneous forms of tokenised money. The taxonomy clarifies how different design choices shape monetary properties, risks, and policy implications, supporting clearer analysis and dialogue across academia, industry, and regulation.
\end{abstract}

\bigskip
\section*{Introduction}
\vspace{0.2cm}

The advent of distributed ledger technology (DLT) has accelerated the emergence of “tokenised money”, digital representations of value that circulate on blockchain networks or similar decentralised infrastructures. Forms of tokenised money range from sovereign issuances to commercial bank tokens and privately issued payment tokens, each with distinct governance and risk profiles. Their characteristics vary depending on factors such as issuer, collateralisation, and technology, which complicates comparisons in a rapidly evolving domain. 

Among the most actively debated forms of tokenised money are stablecoins, deposit tokens, and CBDCs. Stablecoins are blockchain-based instruments pegged to fiat currencies or assets such as gold, typically stabilised through collateral and designed for peer-to-peer payments. Deposit tokens are tokenised commercial bank deposits, backed by bank balance sheets and deposit insurance, offering continuity with today’s banking system while enabling seamless settlements in crypto asset environments. CBDCs are digital central bank liabilities, denominated in sovereign currency. They can be implemented on DLT but do not necessarily have to be. In the context of this article, the taxonomy refers specifically to DLT-based CBDC designs. While all types of tokenised money aspire to function as money, their design, risk profiles, and policy implications differ substantially, making clear classification essential. 

\footnotetext[1]{Hochschule Luzern, Institut für Finanzdienstleistungen Zug IFZ, Suurstoffi 1, 6343 Rotkreuz, Email: \href{mailto:thomas.ankenbrand@hslu.ch}{thomas.ankenbrand@hslu.ch}}

\footnotetext[2]{Hochschule Luzern, Institut für Finanzdienstleistungen Zug IFZ, Suurstoffi 1, 6343 Rotkreuz, Email:          \href{mailto:denis.bieri@hslu.ch}{denis.bieri@hslu.ch}}

\footnotetext[3]{Zürcher Kantonalbank, Zürich-City, Bahnhofstrasse~9, 8001~Zürich, Email: \href{mailto:stefano.ferrazzini@zkb.ch}{stefano.ferrazzini@zkb.ch}}

\footnotetext[4]{ti\&m~AG, Buckhauserstrasse~24, 8048~Zurich, Email: \href{mailto:johannes.hoehener@ti8m.ch}{johannes.hoehener@ti8m.ch}}

\bigskip
\section*{Current market size and potential}
\vspace{0.2cm}

Tokenised forms of money have gained significant attention. According to the Bank for International Settlements~\cite{illes2025} 91 percent of 93 surveyed central banks are engaged in CBDC projects, whether retail focused, wholesale focused, or both, driven by factors such as the decline of cash, the rise of tokenisation, and the growing prominence of stablecoins. To date, however, stablecoins account for nearly all observable volumes, while most deposit tokens and CBDCs remain in the pilot or conceptual phase.

\begin{figure}[!ht]
\centering
\includegraphics[width=0.85\textwidth]{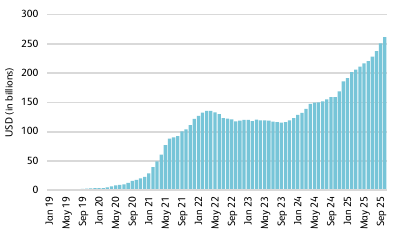}
\caption{Monthly total USD value of stablecoins in circulation (source:~\cite{visa2025a})}
\label{fig:stablecoin}
\end{figure}

As shown in Figure~\ref{fig:stablecoin}, the market capitalisation of stablecoins reached USD 260 billion in October 2025~\cite{visa2025a}. By contrast, no material market volumes exist for commercial bank deposit tokens, and CBDCs are limited to small-scale pilots. This underlines that, despite the existence of multiple designs, stablecoins are currently the only type of tokenised money with significant real-world circulation.

Compared with global reference values, the significance of stablecoins remains limited. Between November 2024 and October 2025, approximately 2 billion transactions were processed using stablecoins~\cite{visa2025b}. In contrast, the global payments industry recorded around 3.4 trillion transactions with a total volume of USD 1.8 quadrillion in 2023~\cite{mckinsey2024}. This illustrates that tokenised money still represents only a marginal share of the overall payments landscape.

Market forecasts nonetheless point to strong growth potential, albeit with considerable uncertainty. J.P. Morgan projects a stablecoin market capitalisation of USD 500 billion by 2028, Standard Chartered places the figure closer to USD 2 trillion by that year, and Bernstein extends the outlook to USD 4 trillion by 2035~\cite{singh2025}. While these figures illustrate the possible trajectory for stablecoins, the potential scale of other forms of tokenised money could be even larger. Based on the financial statements of the world’s 1,000 largest banks, customer deposits collectively amounted to USD 103 trillion in 2024~\cite{tabinsights2025}, while the global money supply (M2) exceeded USD 110 trillion as of mid-2025~\cite{newhedge2025}. If even a fraction of these balances were to migrate into tokenised form, the eventual market size of deposit tokens or retail CBDCs could become substantial, supported by wholesale CBDCs that facilitate tokenised settlement across institutions.

\bigskip
\section*{Why a taxonomy matters}
\vspace{0.2cm}

This scale and diversity underscore the need for a clear taxonomy of tokenised money. Regulatory initiatives such as the European Union’s MiCA framework and the United States’ proposed GENIUS Act introduce stringent requirements for issuance, governance, and reserves, underscoring the urgency of conceptual clarity in distinguishing between different types of tokenised money.

The relevance of such clarity is also evident in Switzerland, where the Swiss Bankers Association~\cite{sba2025} stresses that stablecoins could strengthen the Swiss financial center by improving efficiency, programmability, and innovation, but also warns of risks of disintermediation and dependence on foreign currency solutions if no credible Swiss franc–denominated stablecoin emerges. Only designs fulfilling the “no-questions-asked” principle, backed by high-quality, liquid reserves and issued by supervised intermediaries, are expected to achieve widespread acceptance. Similarly, the TA-SWISS study by Zellweger-Gutknecht et al.~\cite{zellweger2025} emphasises that tokenisation introduces fundamental differences compared to traditional digital money as tokens can be ownership-proximate, decentralised, autonomous, and programmable. These features enable innovative use cases (such as conditional payments in supply chains, insurance, or payment-versus-delivery), but they also entail societal risks regarding privacy, stability, and state control. The study concludes that Switzerland must carefully evaluate the opportunities and risks of stablecoins, tokenised deposits, and CBDCs in order to safeguard financial stability while fostering innovation.

\bigskip
\section*{A taxonomy for tokenised money}
\vspace{0.2cm}

In this environment, a taxonomy of tokenised money is both relevant and necessary. It provides analytical clarity by distinguishing fundamentally different monetary designs, enables comparative evaluation across instruments with divergent risk and governance profiles, and facilitates dialogue between academia, regulators, and industry through a shared conceptual framework. Building on the taxonomy of Ankenbrand et al.~\cite{ankenbrand2024} which distinguished between the token, protocol, and tokenomics dimensions of crypto assets in general, the present framework focuses specifically on tokenised money. A morphological analysis (“Zwicky box”) spanning key design dimensions provides a shared basis for comparison across diverse monetary models. Unlike the broader taxonomy, this specialised framework is motivated by the need to capture the unique monetary, technological, and governance features of instruments that aspire to function as money. It thus offers a common understanding of designs evolving rapidly in a highly dynamic domain, supporting both scholarly analysis and policy discourse.

To operationalise this framework, we set out the key classification dimensions and their possible characteristics. Each dimension reflects a characteristic that differentiates forms of tokenised money, ranging from issuance and governance structures to technological and legal foundations. Figure~\ref{fig:taxonomy} outlines these dimensions, defines their characteristics, and provides examples of real-world projects that illustrate each type.

The dimensions shown in Figure~\ref{fig:taxonomy} are interrelated, as a tokenised money’s unique profile is essentially a combination of the values of each dimension. While the taxonomy provides a systematic framework for classifying the design features of tokenised money, certain critical aspects fall outside its scope.

\newgeometry{left=0.5cm,right=0.5cm,top=0.5cm,bottom=0.5cm}
\begin{landscape}
\thispagestyle{empty}
\begin{figure}
\centering
\includegraphics[height=0.95\textheight]{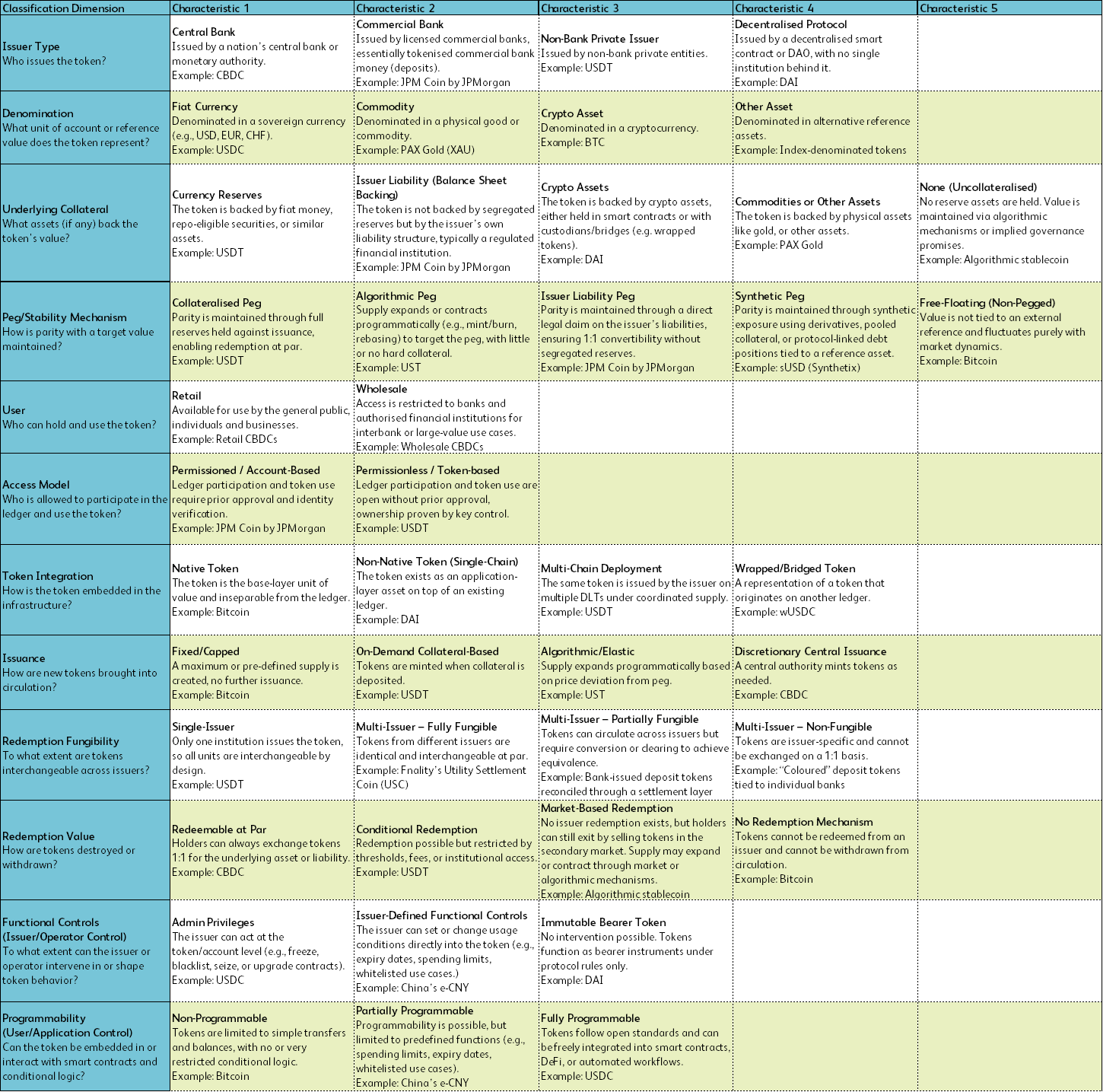}
\caption{Taxonomy for tokenised money}
\label{fig:taxonomy}
\end{figure}
\end{landscape}
\restoregeometry

 These are not design choices per se, but rather consequences or performance indicators of how a tokenised money is structured and embedded in the broader financial system. Among the most important are:

\begin{itemize}
\item Regulatory safeguards: The degree of regulatory protection is a central factor influencing trust and adoption. It encompasses elements such as prudential supervision, backing by regulated balance sheets, and deposit insurance schemes. Importantly, regulatory frameworks differ significantly across jurisdictions, which can lead to varying levels of protection and create disparities in risk perception and market acceptance.
\item Market size and circulation volumes: The actual scale of tokenised money in circulation is a critical performance indicator. At present, volumes remain concentrated in a few types and are modest compared to the global payments landscape.

\item Yield and return potential: The ability of tokenised money to generate interest or rewards is becoming an important factor for competitiveness. Incentive structures such as yield-bearing mechanisms or reward programmes can influence user adoption and shift the role of digital money from a pure means of payment toward a potential savings vehicle. This dynamic also affects the interaction with traditional deposits and broader financial intermediation.

\item Interoperability across issuers: While redemption fungibility within a single issuer is a design dimension in the taxonomy, interoperability across issuers and ecosystems is a strategic challenge. Without common standards or bridging mechanisms, tokenised money risks remaining fragmented into isolated “walled gardens”, limiting its utility for large-scale settlement. Effective interoperability is thus essential for tokenised money to fulfil its potential as a widely adopted medium of exchange.
\end{itemize}

\bigskip
\section*{Examples of tokenised money types}
\vspace{0.2cm}

While the taxonomy is designed to be comprehensive, its value becomes most apparent when applied to specific cases. Mapping well-known forms of tokenised money against the classification dimensions illustrates how different designs cluster, diverge, or overlap. The following sections apply the taxonomy to five representative types, i.e., CBDCs, synthetic CBDCs, fiat-collateralised stablecoins, decentralised protocols, and deposit tokens, to demonstrate how the framework captures both their commonalities and differences:

\subsection*{Central Bank Digital Currencies (CBDCs)}
CBDCs are issued by a \textit{central bank} (Issuer Type) and denominated in the sovereign \textit{fiat currency} (Denomination). They rely on \textit{no separate underlying collateral} since their value derives from sovereign liability (Underlying Collateral) and maintain stability through a \textit{sovereign issuer liability peg} to the national currency (Peg/Stability Mechanism). Access may be \textit{retail} or \textit{wholesale} depending on policy objectives (User) and participation is typically \textit{permissioned and account-based}, requiring identity verification (Access Model). They are often implemented as \textit{native tokens} on permissioned DLT-systems (Token Integration). CBDCs follow \textit{discretionary central issuance} (Issuance) and interchangeability is ensured by the \textit{single issuer} (Redemption Fungibility). Redemption is \textit{at par} into central bank money (Redemption Value). Functional controls may include \textit{admin privileges} or \textit{issuer-defined rules} for compliance and monetary policy tools (Functional Controls). Depending on the design, programmability can range from \textit{non-programmable} (basic transfers) to \textit{partially programmable} (e.g., conditional transfers, expiry dates, whitelisted use cases) (Programmability).

\subsection*{Synthetic Central Bank Digital Currencies (sCBDCs)}
Synthetic CBDCs, sometimes referred to as “reserve-backed tokens” (see, e.g. Goel~\cite{goel2024}), are issued by \textit{commercial banks} or \textit{non-bank private issuers} (Issuer Type), while being fully backed by central bank reserves held in segregated accounts. They are denominated in the sovereign \textit{fiat currency} (e.g., USD, EUR, CHF) (Denomination) and derive their value from \textit{currency reserves}, i.e., central bank money held indirectly through the intermediary (Underlying Collateral). Stability is maintained via \textit{collateralised peg}, ensuring 1:1 convertibility into central bank reserves (Peg/Stability Mechanism). Access is typically \textit{retail}, enabling end users to hold and transfer tokenised representations of central bank money through licensed intermediaries (User). Participation is \textit{permissioned and account-based}, as users must undergo identity verification with the issuing institution (Access Model). The tokens are generally implemented as \textit{non-native} tokens on permissioned or hybrid DLT infrastructures, potentially interoperable with retail payment systems (Token Integration). Issuance is \textit{on-demand collateral-based}, backed by deposits of reserves with the central bank (Issuance). Interoperability across issuers is \textit{multi-issuer} and \textit{fully fungible}, as all tokens represent claims on equivalent central bank reserves and are interchangeable across participating issuers (Redemption Fungibility). Redemption takes place \textit{at par} into sovereign money through the issuing intermediary (Redemption Value). Functional controls are \textit{issuer-defined}, as intermediaries retain the ability to enforce compliance measures such as transaction limits, blacklisting, or KYC enforcement (Functional Controls). Programmability may range from \textit{partially programmable} (e.g., conditional payments, whitelisted purposes) to \textit{fully programmable} when integrated with smart contracts or tokenised payment infrastructures (Programmability).

\subsection*{Fiat-Collateralised Stablecoins}
Fiat-backed stablecoins are typically issued by a \textit{non-bank private issuer} (Issuer Type) and are denominated in a \textit{fiat currency} such as USD (Denomination). Their value is supported by \textit{currency reserves} held with custodians or banks (Underlying Collateral), and stability is maintained through a \textit{collateralised peg} to the reference fiat (Peg/Stability Mechanism). Access is generally \textit{retail}-oriented (User). Participation is typically \textit{permissionless and token-based}, with ownership proven by private key control (Access Model). They are usually implemented as \textit{non-native} tokens on public blockchains (e.g., ERC-20 on Ethereum) and in some cases as \textit{multi-chain deployments} (e.g., USDT across Ethereum, Tron, Solana) (Token Integration). Issuance is \textit{on-demand and collateral-based} when reserves are deposited (Issuance). Interoperability across issuers is generally \textit{single-issuer}, meaning all tokens from one issuer are interchangeable, though fragmentation can occur across chains or competing issuers (Redemption Fungibility). Redemption is typically \textit{at par} but may be \textit{conditional} depending on access rules or fees (Redemption Value). Functional controls often include issuer \textit{admin privileges} such as freezing or blacklisting of addresses (Functional Controls). Most fiat-backed stablecoins are \textit{fully programmable}, integrating seamlessly with smart contracts and DeFi applications (Programmability).

\subsection*{Decentralised Protocols (Algorithmic Stablecoins)}
\textit{Decentralised protocols} issue and manage crypto assets directly through smart contracts without reliance on a central entity (Issuer Type). In the context of tokenised money, they are often referred to as algorithmic stablecoins, which are usually denominated in a \textit{fiat currency} such as USD (Denomination). They typically are \textit{uncollateralised} or only partially backed by \textit{crypto asset} collateral (Underlying Collateral) and maintain value through an \textit{algorithmic peg}, expanding or contracting supply programmatically (Peg/Stability Mechanism). Access is \textit{retail}, with tokens circulating among individuals and businesses (User). Participation is \textit{permissionless and token-based}, with ownership tied to private key control rather than verified identities (Access Model). They operate as \textit{non-native} tokens on public blockchains (e.g., ERC-20 on Ethereum) (Token Integration). Issuance is managed through \textit{algorithmic/elastic} mechanisms, with supply adjusting programmatically to maintain a target value (Issuance). Redemption fungibility is generally \textit{single-issuer} at the protocol level, meaning all units of the token are interchangeable within that system, but there is no cross-issuer fungibility (Redemption Fungibility). Redemption is \textit{market-based}, as value is maintained through trading dynamics rather than issuer-backed guarantees (Redemption Value). They typically lack issuer-admin controls, functioning instead as \textit{immutable bearer tokens} governed solely by protocol rules (Functional Controls). Most are \textit{fully programmable} and integrate directly into smart contracts and DeFi applications (Programmability).

\subsection*{Deposit Tokens (Commercial Bank-Issued Tokens)}
Deposit tokens are issued by \textit{commercial banks} (Issuer Type) and are typically denominated in \textit{fiat currencies} such as USD, EUR, or CHF (Denomination). Their value is supported by \textit{issuer liability (balance-sheet backing)}, meaning they rely on the regulated deposit framework rather than segregated reserves (Underlying Collateral). They maintain parity through an \textit{issuer liability peg}, whereby the claim on the bank’s liabilities ensures 1:1 convertibility with fiat currency (Peg/Stability Mechanism). Access may be \textit{retail} or \textit{wholesale} depending on design (User). Participation is usually \textit{permissioned and account-based}, though retail variants may incorporate token-based (bearer-like) features (Access Model) and the tokens may be implemented as either \textit{native} tokens on a dedicated ledger or as \textit{non-native} tokens on existing platforms (Token Integration). Issuance follows \textit{discretionary central issuance} by the bank (Issuance), and interchangeability across issuers depends on the issuance model: \textit{single-issuer} tokens are fungible by design within that bank, while \textit{multi-issuer} models may be fully \textit{fungible}, \textit{partially fungible}, or \textit{non-fungible} across different banks (Redemption Fungibility). Redemption is \textit{at par} into deposits (Redemption Value). Functional oversight is strong, with \textit{admin privileges} ensuring compliance and risk management (Functional Controls). Depending on system design, they may be \textit{non-programmable} or \textit{partially programmable} (e.g., conditional settlement, whitelisted use cases) (Programmability).

\bigskip
\section*{Use cases of tokenised money}
\vspace{0.2cm}

Beyond their technical design, the relevance of tokenised money is ultimately determined by its practical applications. Different forms support diverse payment and settlement needs across retail, wholesale, and decentralised environments. In retail payments, tokenised money can enable instant peer-to-peer transfers, micro-payments, and pay-per-use or machine-to-machine (M2M) transactions within digital and IoT ecosystems. Retail CBDCs may serve as a secure, inclusive public payment option, while private stablecoins already facilitate low-cost, global, and programmable transfers in crypto asset markets. In wholesale and institutional finance, tokenised money supports faster, safer, and more efficient settlement of interbank, corporate, and capital market transactions. Wholesale CBDCs and deposit tokens enable atomic settlement on distributed ledgers, reducing counterparty and operational risks. Synthetic CBDCs combine central-bank-backed safety with private-sector distribution, extending similar efficiencies to institutional payments. Tokenised money also enables programmable, automated (and conditional) transactions, where smart contracts trigger payments once predefined conditions are met. This supports automated trade and supply-chain finance, insurance payouts, and other conditional settlements (delivery-versus-payment). Across all types, tokenised money has strong potential for cross-border payments, offering faster, cheaper, and more transparent international transfers. Together, these use cases illustrate tokenised money’s role as a programmable, interoperable, and efficient medium of exchange, bridging traditional and digital financial systems.

\bigskip
\section*{Summary and conclusion}
\vspace{0.2cm}

Tokenised money is not a single innovation but a spectrum of designs with distinct issuers, collateral structures, and governance arrangements. The taxonomy presented structures this diversity across twelve classification dimensions, offering a systematic way to compare forms such as CBDCs, fiat-collateralised stablecoins, algorithmic stablecoins, and deposit tokens. This clarity is urgent given the significant disparity between current market volumes and their theoretical potential. The framework highlights how stability is achieved, where risks concentrate, and which governance and technological arrangements underpin trust. Just as importantly, the value of tokenised money lies in its applications and adoption, which are manifold, ranging from retail and wholesale payments to settlement, micro-payments, machine-to-machine transactions, and automated conditional processes. Beyond academic inquiry, the taxonomy provides policymakers and industry with a common language for assessing design trade-offs, regulatory implications, and market developments. As tokenisation advances, it will remain a practical tool for analysing emerging projects and understanding the evolving landscape of tokenised money. In Switzerland, this development is now also reflected in regulatory efforts: in October 2025, the Federal Council opened a consultation on amendments to the Financial Institutions Act. The initiative aims to establish a clear legal framework for the issuance of stablecoins and other forms of tokenised money, introducing new licence categories for “payment instrument institutions” and “crypto-institutions” to foster innovation while safeguarding financial stability and consumer protection~\cite{sif2025}.

\bigskip
\section*{Acknowledgements}
\vspace{0.2cm}

The work was carried out within the IFZ FinTech Program and was financially supported by various industry partners and the Lucerne University of Applied Sciences and Arts. The authors gratefully acknowledge the constructive input and practical insights provided through the collaboration with the program’s industry partners.

\bigskip
\section*{Note and Disclaimer}
\vspace{0.2cm}

The proposed taxonomy should be understood as a draft classification framework for tokenised money, intended to stimulate further discussion and refinement. Feedback and suggestions are warmly welcomed. The taxonomy and paper have been prepared to provide general information. Nothing in this paper constitutes a recommendation for the purchase or sale of any financial instrument. In addition, the paper includes information obtained from sources believed to be reliable, but the authors do not warrant their completeness or accuracy. This also includes the outputs of AI tools, like ChatGPT or DeepL, which were situationally used in the preparation of this paper.

\bigskip

\bibliographystyle{amsplain}
\bibliography{references}

\end{document}